\documentclass[10pt, conference, compsocconf]{IEEEtran}

\def\IEEEbibitemsep{0pt plus .3pt}

\usepackage{cite}
\usepackage{amsmath,amssymb,amsfonts}

\usepackage{graphicx}
\graphicspath{ {images/} }
\usepackage[caption=false]{subfig}

\usepackage[outdir=./]{epstopdf}
\usepackage{textcomp}
\usepackage{xcolor}
\usepackage{url}

\let\labelindent\relax

\usepackage{enumitem}

\usepackage{algorithm}
\usepackage{algorithmicx}
\usepackage[noend]{algpseudocode}

\usepackage[utf8]{inputenc}
\usepackage[english]{babel}
\usepackage{lipsum}

\usepackage{amsthm}
 
\theoremstyle{definition}
\newtheorem{definition}{Definition}

\begin{document}
%
\title{ Event-based  Detection of Changes in IaaS Performance Signatures \vspace{-6mm}}



\author{\IEEEauthorblockN{Sheik Mohammad Mostakim Fattah and Athman Bouguettaya}
\IEEEauthorblockA{School of Computer Science, University of Sydney, Australia\\
Email: \{sheik.fattah, athman.bouguettaya\}@sydney.edu.au} \vspace{-6mm}
}

\makeatletter
\def\ps@IEEEtitlepagestyle{
  \def\@oddfoot{\mycopyrightnotice}
  \def\@evenfoot{}
}

\def\mycopyrightnotice{
  {\footnotesize
  \begin{minipage}{\textwidth}
  \centering
  Copyright~\copyright~2020 IEEE.  Personal use of this material is permitted.  Permission from IEEE must be obtained for all other uses, in any current or future media, including reprinting/republishing this material for advertising or promotional purposes, creating new collective works, for resale or redistribution to servers or lists, or reuse of any copyrighted component of this work in other works.
  \end{minipage}
  }
}


%


\maketitle

\begin{abstract}
We propose a novel ECA approach to manage changes in \textit{IaaS performance signatures}. The proposed approach relies on the detection of \textit{anomalous} performance behavior in the context of IaaS performance signatures. A novel anomaly-based event detection technique is proposed. It utilizes the experience of free trial users to detect potential changes in IaaS performance signatures. A signature change detection technique is proposed using the cumulative sum control chart analysis. Additionally, a self-adjustment method is introduced to improve the accuracy of the proposed approach. A set of experiments based on real-world datasets are conducted to show the effectiveness of the proposed approach.

\end{abstract}


%
\IEEEpeerreviewmaketitle

\section{Introduction}

Infrastructure-as-a-Service (IaaS) is a key service delivery model in the cloud market \cite{iosup2014iaas}. Various computational resources such as CPU, memory, and storage are offered through IaaS models in the form of Virtual Machines (VMs). IaaS cloud enables an easier, faster, and cost-effective way to migrate and manage an organization's in-house IT infrastructure in the cloud \cite{iosup2011performance}. Large business organizations typically prefer to utilize IaaS services on a \textit{long-term} basis \cite{mistry2018metaheuristic}. Most IaaS providers offer significant discounts on long-term subscriptions (e.g., 1 to 3 years) in the cloud market. For example, Microsoft Azure offers up to 72\% discounts on long-term subscriptions.

IaaS service \textit{selection} for a long-term period is an important \textit{business decision} for many organizations due to economic reasons \cite{fattah2018cp}. The performance of IaaS services is a major concern during long-term selection \cite{mistry2018metaheuristic}. Selecting a service that may perform poorly in the future, may lead to an inevitable \textit{loss of productivity} for an organization. The performance of an IaaS service is typically measured in terms of Quality of Services (QoS) attributes such as CPU execution time, disk read/write throughput, and latency. QoS attributes help a consumer to determine the \textit{best performing} services from a large number of functionally similar services \cite{chaki2020fine}. The long-term QoS-aware service selection is therefore defined as the similarity matching between the consumer's long-term QoS requirements and the expected long-term performance of IaaS services \cite{fattah2019long}. 

The knowledge of the IaaS services' performance is paramount in determining which ones are the best fit for the consumers' required QoS  \cite{iosup2011performance}. Despite that, IaaS providers typically reveal very \textit{limited} performance information in their advertisements due to \textit{market competition} and \textit{business secrecy} \cite{wenmin2011history}. For instance, most IaaS advertisements do not contain information about actual vCPU (virtual CPU) speed, memory bandwidth, or VM startup time. The performance of a VM may change over time given the dynamic nature of the cloud environment \cite{iosup2011performance}. As a result, advertised performance information may not reflect the actual service performance for a particular provisioning time. For example, a consumer may want to utilize some VMs in December where the advertised performance is measured in June. In such a case, the advertised information is not useful for the selection in December. Additionally, the advertised performance information may not be \textit{helpful} to understand service performance due to the lack of detailed information \cite{wenmin2011history}. For instance, Amazon EC2 instances have different types of virtual CPUs (vCPUs). According to the EC2 advertisements, each vCPU is either a thread of  an Intel Xeon core or AWS Graviton processor\footnote{https://aws.amazon.com/ec2/instance-types/}. Estimating the performance of the vCPU from such incomplete information is difficult. Therefore, the lack of detailed and complete performance information makes the long-term selection challenging \cite{fattah2019long}. 

Effective utilization of \textit{free trials} offered by IaaS providers is a unique way to deal with the limited performance information for the long-term selection \cite{wang2018testing}. A long-term IaaS selection framework is proposed in \cite{fattah2019long}. It leverages short-term trials to discover the unknown performance information of an IaaS service \cite{fattah2019long}. The framework introduces an equivalence partitioning-based strategy that maps a consumer's long-term workloads into the free short-term trial periods to discover long-term performance. However, free trial experiences do not provide adequate information to make the best service selection for a long-term period. The key reason is that the performance of IaaS services changes \textit{periodically} due to the multi-tenant nature of the cloud \cite{iosup2011performance}. The observed performance in a trial in one month may change if the trial is performed in a different month. Therefore, making a long-term commitment based on only short trials does not always lead to the best service selection \cite{fattah2019long}.

\textit{IaaS performance signatures} offer an effective alternative to deal with the unknown service performance variability for the long-term selection \cite{mi2008analysis,fattah2020icws}. An IaaS performance signature represents the \textit{expected} performance behavior of an IaaS service over a long period of time. For instance, a signature of a VM may indicate that its response time is expected to increase by 10\% on January than the response time in December. A consumer's trial experience of a service and its corresponding signature can be utilized together to make a better selection for the long-term period. A signature-based IaaS selection approach is proposed in \cite{fattah2020icws}. The proposed approach generates signatures using the experience of past trial users. However, the proposed approach does not consider the \textit{dynamic} nature of performance signatures. 


IaaS performance signatures are dynamic in nature and may need to be re-evaluated over a long period of time for a number of reasons \cite{mi2008analysis}. For instance, a provider may upgrade its infrastructure or change its multi-tenant management policy resulting in change of service performance \cite{leitner2016patterns}. In such a case, it is important to detect the change of IaaS performance \textit{as early as} possible to make sure its signature reflects the \textit{current} performance behavior of the service. \textit{We focus on the detection of change in IaaS performance as represented by its signature}. In this case, the IaaS performance signature may need to be updated to be representative of the new performance profile of the service. We propose a novel \textit{Event-Condition-Action} (ECA) approach to mange changes in IaaS performance signatures. The ECA model is a simple yet powerful tool that has been extensively used in databases, cognitive computing, and semantic web. In the ECA model, when an \textit{event} is detected, a \textit{condition} is checked, and a resulting \textit{action} is executed \cite{liu2009encyclopedia}. 

We identify two key challenges in IaaS performance signature change detection. The first challenge is determining the \textit{threshold} which would trigger testing whether the present signature needs to be \textit{re-evaluated} \cite{veeravalli2014quickest}. Certain changes in performance of a service may not necessitate a change in its signature.  For instance, a major failure of computing infrastructure may negatively impact the performance of an IaaS service at a point in time without necessarily indicating a long term change in the performance behavior.  Therefore, the challenge is to \textit{accurately} identify situations where there is a high likelihood of long-term changes in performance, thus requiring a re-evaluation of the signature. The second challenge is ascertaining whether the re-evaluation of the signature was warranted. As predicting whether a change of performance warrants a re-evaluation of the signature, is probabilistic in nature, there is a need to ascertain that it was the correct course of action. The challenge is to identify factors that would evaluate the accuracy of that re-evaluation. 

We propose a set of techniques that rely on the detection of anomalous performance behavior in the context of IaaS performance signatures. In particular, the proposed approach consists of two main parts: a) an \textit{anomaly-based event detection} technique that determines when to trigger the re-evaluation of a signature, and  b) \textit{a signature change detection} method that leverages time series change detection techniques to re-evaluate existing IaaS performance signatures. In addition, we introduce a self-adjustment method to improve the performance of the proposed ECA approach using a feedback loop from the outcome of the signature change detection. In summary, we propose a novel framework for the detection of accurate changes in IaaS performance signatures.  Accuracy is achieved over time through continuous testing of the re-evaluated signatures which may lead to either (1) confirming the previous signature changes, or (2) invalidating the previous signature changes.

\section{IaaS Performance Signature}

We discuss how to represent the performance signature of an IaaS service in this section. The word ``signature'' typically refers to the unique characteristics or behavior of an object, entity, or piece of information. The concept of the signature has been widely utilized in a number of domains such as cryptography, security, computing, and mathematics. For instance, signatures of various application performance are utilized for resource capacity planning and performance anomaly detection \cite{mi2008analysis}. Intrusion Detection Systems (IDS) leverage signature-based malware detection techniques to enable quick detection of security threats. Digital signatures are commonly used to verify the authenticity and integrity of digital messages or documents. 



We represent the signature of an IaaS service based on its \textit{relative} performance changes over time, i.e., how much a service's performance may increase or decrease in one time compared to another time. For example, the signature of a VM may inform that its response time is expected to increase by 5\% on weekend nights than regular weekdays. Note that, the signature does not tell the exact performance of a service. Therefore, a consumer is unable to select a service based on only its signature. Instead, the consumer needs to perform the trial with its application workloads and utilize the trial experience and the IaaS signature to estimate the long-term service performance \cite{fattah2020icws}. 

\begin{definition} {IaaS Performance Signature}: An IaaS performance signature is a temporal representation of relative performance changes of an IaaS service over a long period.

\end{definition}

The signature is represented by a set of QoS parameters that are relevant to the service. The \textit{relevant} QoS attributes are defined by the most important QoS attributes to measure the performance of a particular type of IaaS service \cite{fattah2020icws}. For example, data read/write throughput, and disk latency are the key QoS attributes for virtual storage services.



We denote the signature of a service as $S=\{S_1, S_2,...S_n\}$, where $n$ is the number of QoS attributes in the signature. Each $S_i$ corresponds to a QoS attribute. Each $S_i$ denotes a time series for $t$ period which is represented as $S_i = \{s_{i1},s_{i2},......s_{it}\}$. Here, $s_it$ is the relative performance of the provider at the time $t$ for a particular QoS attribute. We use the following representation to denote a signature:

\scriptsize
\begin{gather}
 S =
  \begin{bmatrix}
   s_{11} & s_{12} & .. & s_{1t} \\
   s_{21} & s_{22} & .. & s_{2t}  \\
  s_{31} & s_{13} & .. & s_{3t}  \\
   .. & .. & ... \\
   s_{n1} & s_{n2} & .. & s_{nt}  \\
   \end{bmatrix}
   \label{eqn:signature}
   \vspace{-.2cm}
\end{gather}

\normalsize where each row corresponds to the QoS signature of $Q_i$ and each column represents a timestamp $t$. From the equation \ref{eqn:signature}, we see that a signature may include several QoS attributes. However, we focus only one QoS attribute in this work, i.e., throughput of an IaaS service for simplicity. As a result, the signature in this work is considered two-dimensional. We may extend this work in future to support more than two dimension of IaaS performance signatures.

\section{Generation of IaaS Performance Signatures}

\label{sec:sig}

It is important to note that, the past trial users may not want to share their experience publicly to protect their privacy, security, and the conflict of interests with the provider \cite{zhu2015privacy}. However, they may share their trial experience with a \textit{Trusted} Non-Profit Organization (TNPO) for a limited period to help new consumers in the selection \cite{van2012trusted}. Examples of such TNPOs are available in public sectors where privacy-sensitive information about individuals needs to be shared to deliver better services. For instance, health research institutes often collect data about individual patients to improve health services. TNPOs are responsible for data \textit{integration} and \textit{distribution} of collective knowledge without revealing individual's privacy-sensitive information. 

\begin{figure}[b]
 \vspace{-.4cm}
    \centering
    \includegraphics[width=.48\textwidth]{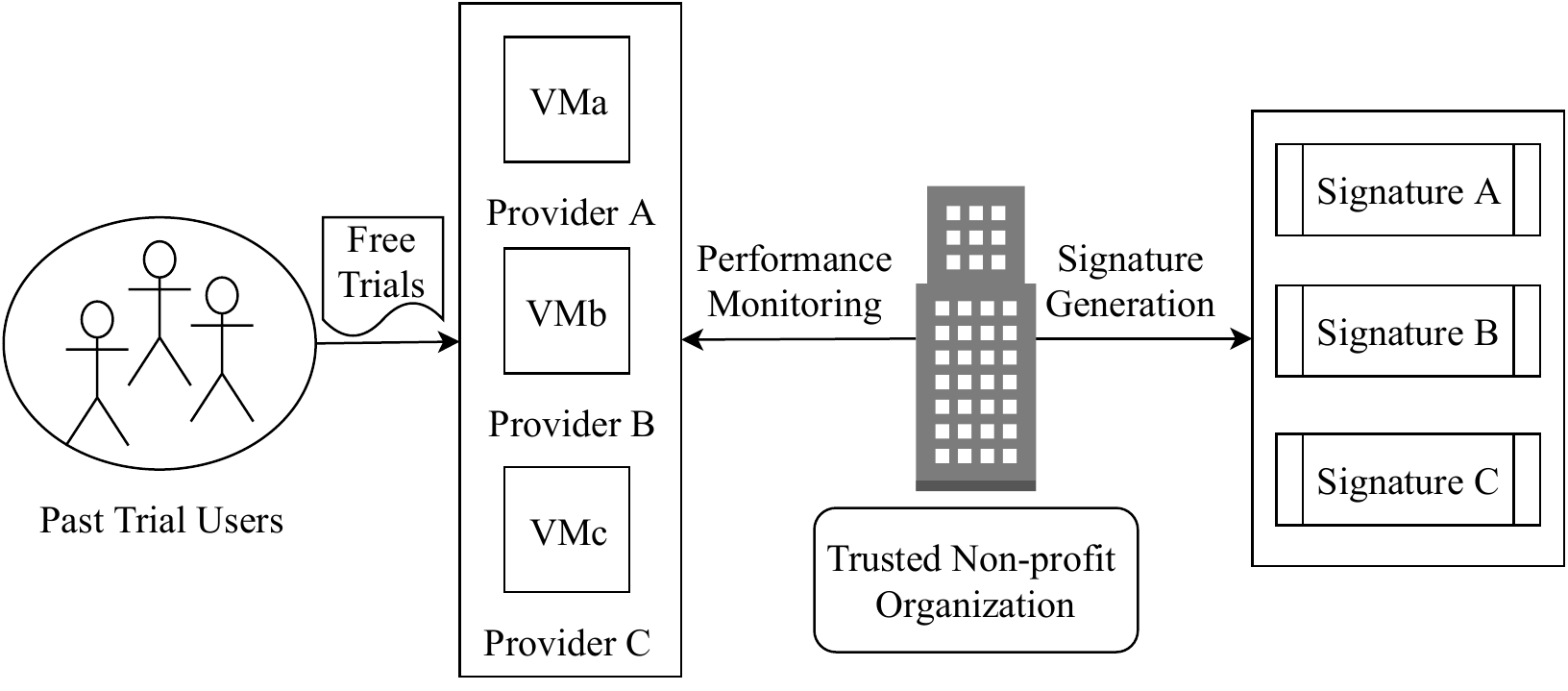}
    \caption{IaaS performance signature generation}
    \label{fig:tnpo}
\end{figure}

\textit{We assume that the past trial users who have utilized some IaaS services share their experience with a TNPO for a limited period of time.} The TNPO generates IaaS performance signatures based on the aggregated experience of past trial users and deletes the users' data afterward. Let us assume that there are three IaaS providers ($A$, $B$, and $C$) who offer three VMs ($VM_a$, $VM_b$, and $VM_c$) with similar configurations (e.g., resource capacity, location) for free short-term trials as shown in Fig. \ref{fig:tnpo}. There are past users who utilized the VMs to find the performance over different periods of time. The trial users do not want to share their trial experience publicly. However, each trial user shares its experience with a TNPO for a short period. The TNPO generates the signature to identify the long-term performance variability of each VM. The TNPO has to delete users' experience once the signatures are computed. A signature provides an aggregated view of a VM's long-term performance variability. It is not possible to derive individual trial experience from the signature. As a result, the TNPO does not violate the privacy of past trial users.

We create IaaS performance signatures in a way that requires less detailed performance information about the service performance and the past trial users and yet useful enough to make a long-term selection. We apply a \textit{normalized averaging} method \cite{fattah2019long} to generate the signature based on the experience of past trial users. Let us assume that $k$ number of past trial users share their observed trial performance $Q_{k}$ over the period $T$ for a service. Here, $Q_{k}$ refers to the performance observed by the $k$th consumer for the QoS attribute $Q$ over the period $T$. We denote $Q_{k}$ as $Q_{k}=\{q_{1k}, q_{2k},.., q_{tk}\}$. The following steps are performed to generate the signature for the QoS attribute $Q$:

\begin{enumerate}[itemsep=0ex, leftmargin=*]
    \item For a QoS attribute $Q$, the performance observed by the trial users is collected over time $T$.
    \item At each timestamp $t \in T$, the average performance observed by $k$ number of consumers is measured for $Q$. The average performance is denoted by $\overline{Q_{k}}$.
    
    \item Each $\overline{Q_{k}}$ is normalized based on its standard deviation $\sigma (\overline{Q_{k}})$. The normalized QoS time series is considered as the IaaS performance signature $S$ over the period $T$. 
\end{enumerate}

The value of $s_nt$ at any $t$ represents the relative QoS performance compare to any other time $t'$ in Equation \ref{eqn:signature}. This representation of the signature offers two benefits. First, the use of signature becomes easier once a consumer has utilized free trials based on its workloads. The performance for any other time can be found by comparing the ratio between the trial month and other times. Second, signatures can be stored and updated easily over time as it does not require storing detailed information. The signature mainly reflects \textit{substantial changes} in the performance over a long period. The effect of the signature should be visible by most consumers in the trial period unless the provider utilizes an isolated environment.

\section{Proposed ECA Approach}

We apply an ECA approach to manage changes in IaaS signatures. The ECA model is especially useful when an action needs to be performed based on a condition that needs to be satisfied. According to the ECA model, an event determines when to trigger an action, the condition defines how to evaluate the event, and the action sets the execution plan in response to the event. An event is typically a special indicator that informs a system that an action may need to be performed. An example of events in security software could be defined as the deletion of a large number of files at once. The security software may start the evaluation of the event, i.e., the deletion of a large number of files to find out whether it is a result of a security attack or a user action. 

Anomalous performance behavior is a potential indicator of IaaS performance change \cite{mi2008analysis}. An anomalous performance behavior is the \textit{deviation} from the expected performance of an IaaS service. The expected performance is represented by its performance signature. Performance anomalies are typically common in cloud environment \cite{ibidunmoye2015performance}. Anomalies may occur due to unexpected events faced by the IaaS provider such as a sudden increase in the workload of the physical system, power failure, or natural disasters. As a result, experiencing performance anomalies in the free trial period may be normal in cloud. However, the frequent occurrence of performance anomalies in the free trial period may indicate changes in IaaS performance, thus requiring a re-evaluation of the existing signature. \cite{mi2008analysis}. Therefore, we define the event for the IaaS performance change detection based on the frequent occurrence of performance anomalies.

\begin{definition} {\textit{Event}:} An event is the frequent occurrence of performance anomalies that are experienced by the free trial users within a fixed period of time.
\end{definition}

The frequency is initially defined as an arbitrary number or threshold $f$ which can be adjusted in the self-adjustment step. Once an event is detected, it needs to be evaluated to detect whether the signature has been changed. If the event satisfies the condition, the signature needs to be updated. The condition and action are defined as follows:

\begin{definition} {\textit{Condition}:} The condition is the process of testing an event to ascertain changes in IaaS performance.
\end{definition}

\begin{definition} {\textit{Action}:}
The action is the process of updating the present IaaS performance signature to reflect the changes in the IaaS performance.

\end{definition}

We utilize the above three definitions as the basis for the proposed ECA approach. In the following sections, we describe the three parts of the proposed approach: a) an anomaly-based event detection, b) a signature change detection and signature update (condition and action respectively), and c) a self-adjustment method to improve the accuracy of the proposed approach.

\section{Anomaly-based Event Detection}

We utilize the free trial experience and the existing IaaS performance signatures to detect performance anomalies. The events are detected based on the performance anomalies. First, we measure the similarity between the trial experience of a consumer and the signatures to detect performance anomalies \cite{ibidunmoye2015performance}. When a user's trial experience is similar to the current signature, the signature is considered to be representative of the expected service performance. When the trial experience does not exhibit similar performance behavior as represented by its signature, we consider it as an anomalous performance behavior. The signature represents the relative performance behavior as a time series. As a result, the shape of the time series needs to be considered to measure the similarity rather than the value of each data point in the signature time series. There are numerous approaches in the existing literature to measure time series similarity based on the shape such as Pearson Correlation Coefficients (PCC), Euclidean Distance (ED), Spearman Correlation (SC), Cosine Similarity (CS), Symbolic Aggregate Approximation (SAX), and Dynamic Time Warping (DTW). Each of these methods may be applied to measure the similarity between the trial experience and the signatures. We briefly discuss how to apply the PCC, CC, and ED for the similarity measure to detect performance anomalies in the context of IaaS performance signatures.



Let us denote the trial experience of a user by $E_Q$ where $E_Q$ denotes the performance of an IaaS services in the free trial period  $T_f$. Here, $T_f << T$, i.e., the free trial period is significantly less than the required provisioning time $T$. We represents $E_Q$ as a time series $E_Q = \{q_1, q_2, ...q_{n}\}$ where $n$ is the number of timestamps in $t$. $E_Q$ needs to be normalized before measuring the similarity with an IaaS performance signature. Let $E'_Q$ is the normalized trial performance where the normalization is performed based on its standard deviation. We denote $E'_Q$ as $E'_Q = \{q'_1, q'_2, ...q'_{n}\}$. Let the signature of an IaaS service for the trial period $T_f$ is $S_Q$ for the QoS attribute Q where $S_Q = \{s_1, s_2, s_3,...s_n\}$. The similarity between the normalized trial experience $(E'_Q)$ and the signature of a service during the trial period $(S_Q)$ using the Euclidean distance $(S(E'_Q, S_Q)^{ED})$ is computed by the following equation:




\begin{equation}
    S(E'_Q, S_Q)^{ED} = \sqrt{\sum_{t=1}^n (s_t - q'_t) }
\end{equation}

\normalsize

Similarly, the similarity measure using the Pearson Correlation Coefficients is computed using the following equation: 

\begin{equation}
    S(E_Q^N, S_Q)^{PCC} = \frac{\sum_{t=1}^n (s_t - \Bar{s}) (q'_t - \Bar{q'}) }{\sqrt{ (s_t - \Bar{s})^2 } \sqrt{(q'_t - \Bar{q})^2}}
\end{equation}

\normalsize

where $\Bar{q'}$ and $\Bar{s}$ is the mean value of $q'$ and $s$ within the trial period $T_f$. The cosine similarity of the trial experience is measured by the following equation: 

\begin{equation}
    S(E_Q^N, S_Q)^{CS} = \cos{\theta} = \frac{\sum_{t=1}^n s_t  q'_t }{ \sqrt{\sum_{t=1}^n (s_i)^2 } \sqrt{\sum_{t=1}^n (q')^2_i}  }
\end{equation}

\normalsize

Each of the above equations provides us with a similarity value between the trial experience and the corresponding IaaS performance signature. In the case of the euclidean distance, the lower the distance is the higher the similarity. 

A similarity threshold needs to be defined to determine how much deviation of the performance from the signature should be considered as the performance anomaly. We define a similarity threshold $S_{thresh}$ for each technique. The threshold is used to distinguish between the normal performance behavior and performance anomalies. The initial threshold is defined during the signature generation process based on the experience of the past trial users' experience. Let us assume that there are $N$ number of past trial users. The experience of the past trial users is denoted by $E_P = \{E_1, E_2, ...E_N\} $. The initial similarity threshold $T_S$ for anomaly detection is defined as follows:


\begin{equation}
    T_S = \min_{i=1}^N{S(E_i, S_Q)}^{M}
\end{equation}

\normalsize

where $M$ denotes the similarity measure method, i.e., PCC, ED, or CS. The threshold for different similarity measure technique can be different. When a new user performs trial if the user's observed performance has a similarity lower than the $T_S$, we consider it as anomalous performance behavior of the service. The value of the similarity threshold $S_{thresh}$ is adjusted based on the experiments. 

The event for signature change detection is defined as the frequent occurrence of performance anomalies within a fixed period of time as mentioned earlier. Therefore, we define an anomaly threshold for the event detection and denote as $F_{thresh}$ which represents the minimum number of occurrence of the performance anomalies within a period of time $T_f$. The value of $T_f$ is the length of the free trial period. We assume that each provider offers the same length of free trial without the loss of generality. The value of $F_{thresh}$ can be initially defined as the number of past trial users within each $T_f$ period who have the minimum similarity between their experience and the corresponding signature. For example, if there are 5 past trial users who have the minimum similarity $T_S$ with the present signature, then $F_{thresh}$ is initialized as $5$. In such a case, the number of past trial users that have the minimum similarity during the signature generation process is considered as the usual number of performance anomalies within $T_f$ period. When the number of performance anomalies crosses $F_{thresh}$, we consider it as an event that needs to be evaluated for the signature change detection. We update the value of $F_{thresh}$ over time to detect the signature change effectively in the self-adjustment step based on the experiments.

\section{Signature Change Detection}

An event indicates that a signature may need to be re-evaluated. When an event is detected, the present signature needs to be tested to evaluate the event. This testing is the condition part of the proposed ECA approach. The main concern in the signature change detection is to differentiate between the performance anomalies and performance changes. This is similar to the signature processing domain, where the noise is a major concern for signal change detection. For instance, a voice recognition program has to differentiate between the noises in the environment and the voice of new persons. There exist a number of approaches for change detection in a signal or time series based on supervised, semi-supervised, or unsupervised methods \cite{aminikhanghahi2017survey}. We choose an unsupervised method called CUSUM which is a sequential analysis technique for small change detection in a time series. The CUSUM control chart is a simple and effective technique that is used in several areas such as signal processing, image processing, and intrusion detection in computer networks and security systems \cite{veeravalli2014quickest}.

A CUSUM control chart monitors the deviation of the individual or a group of samples from a target mean. Let us assume that the observation of a process $P$ has the following sequence $x_1, x_2, ...x_n$ with an estimated average of $m_x$ and standard deviation $s_x$. The upper limit and the lower limit of the cumulative sum is defined by the following equations: 


\begin{equation}
    UL_i= 
    \begin{cases}
        \max{(0, UL_{i-1}+x_i-m_x- \frac{1}{2} n s_x) }                ,& i \geq 1\\
        0,              & i = 1
    \end{cases}
    \label{eqn:ul}
\end{equation}

\vspace{-4mm}

\begin{equation}
     LL_i= 
    \begin{cases}
        \min{(0, LL_{i-1}+x_i-m_x + \frac{1}{2} n s_x) }                ,& i \geq 1\\
        0,              & i = 1
    \end{cases}  
    \label{eqn:ll}
\end{equation}

\normalsize

where $UL_i$ is the upper limit, $LL_i$ is the lower limit, $n$ is the \textit{minimum detectable shift} from the target mean. The process $P$ is considered in violation of CUSUM criteria at the sample $x_i$ if it obeys $UL_i > cs_x $ or $LL_i < -cs_x $ where $c$ represents the control limit. The value $c$ is adjustable and represents the number of standard deviations that the upper and lower cumulative sums are allowed to drift from the target mean. 

Once an event is detected within a period of time $T_f$, we recompute a new signature $(S^N)$ based on the trial experience of all the users within that period of time using the signature generation technique described in section \ref{sec:sig}. The CUSUM control chart is applied to the new signature based on the equation \ref{eqn:ul} and \ref{eqn:ll}. The target mean $m_x$ and the standard deviation $s_x$ is set based on the existing signature $S$ within $T_f$ period. The value of $c$ and $n$ is set based on the standard practice of CUSUM that is $s_x$ and $5s_x$ respectively. Once we detect the change in the IaaS performance signature within a time window of $T_f$, the existing part of the signature is replaced by the new signature. 

\section{Self-adjustment of the ECA Approach}

When an event is detected and evaluated based on the proposed signature change detection technique, the outcome will be either a true positive or false positive. A true positive implies that the signature needs to be updated. A false positive indicates that the signature does not need to be updated. The number of false positives can be reduced by adjusting the similarity threshold $S_{thresh}$ for the anomaly detection and the anomaly threshold $F_{thresh}$ for the event detection. For example, let us assume that the anomaly threshold for the event detection is set to 5 performance anomalies for a one-month trial period. If the TNPO detects 5 anomalies in every month and the outcome is a true positive, then the anomaly threshold should have been reduced earlier to detect the change in IaaS performance. Similarly, if the outcome is false positive every time, we need to increase the anomaly threshold for event detection. We apply a self-adjustment method using a feedback loop from the outcome of the signature change detection to the event detection to change the anomaly threshold.

\begin{figure}
    \centering
    \includegraphics[width=.4\textwidth]{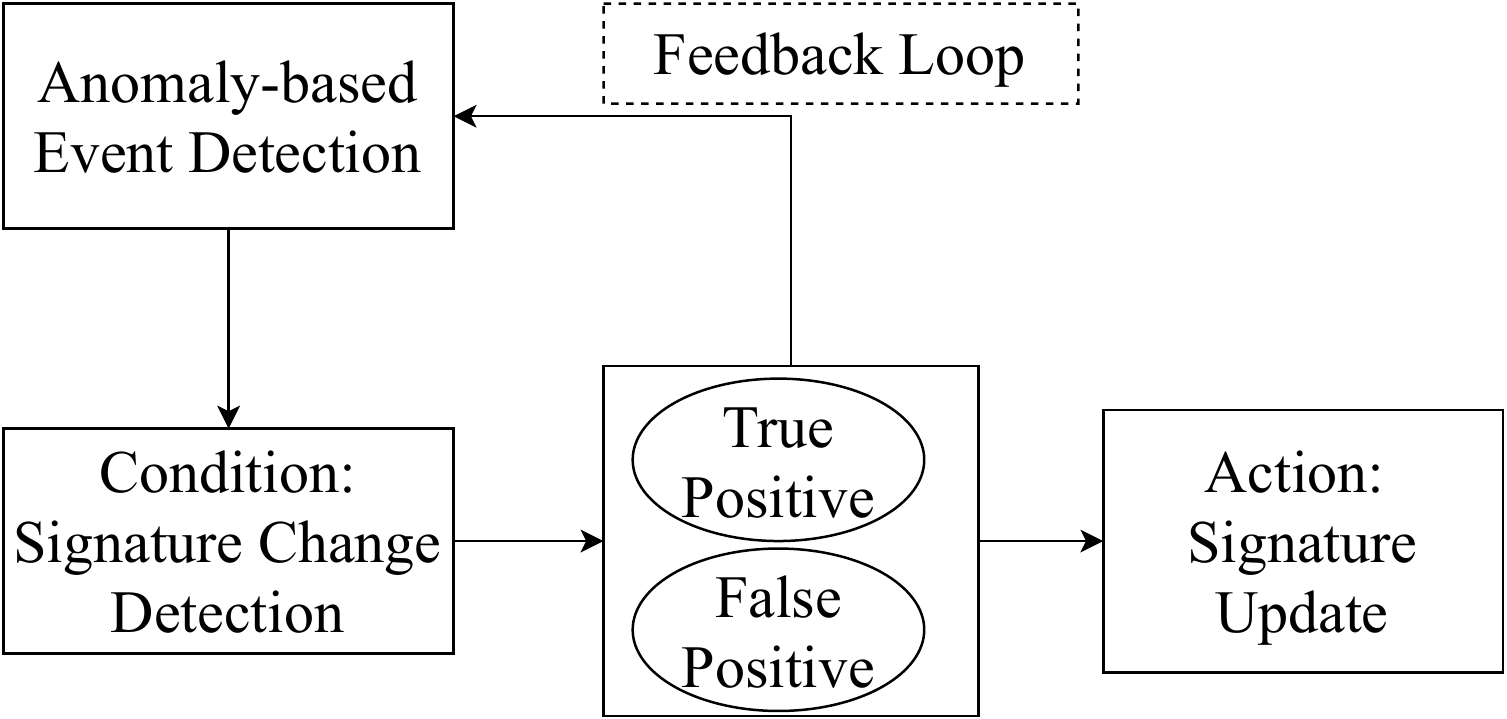}
    \caption{Self-adjustment of the ECA approach}
    \label{fig:feedback}
    \vspace{-.6cm}
\end{figure}

Fig. \ref{fig:feedback} shows the proposed self-adjustment method using a feedback loop. The outcome of the condition checking is fed to the anomaly-based event detection module. When the number of true positives or false positives  within a predefined period of time $T'$ exceeds a predefined threshold $Z$, the event detection module updates the frequency threshold $F_{thresh}$. The value of $Z$ and $T'$ is set by the TNPO. The frequency threshold is updated linearly based on the change detection outcome. when the outcome of signature change detection exceeds the true positive threshold then $F_{thresh}$ is incremented by one. If the outcome exceeds the false positive threshold, then $F_{thresh}$ is decremented by one.

\section{Experiments and Results}

A series of experiments are conducted to evaluate the proposed ECA approach. We identify two key attributes: a) the number of false positives, and b) the change detection delay to evaluate the proposed approach. 

\subsection{Experiment Setup}

Finding real-world workload traces and performance datasets for a long-term period is very challenging. Thus, we utilize the publicly available workload traces and performance data to mimic the long-term cloud environment. We use the Eucalyptus IaaS workload to generate the trial workloads of different consumers \footnote{\url{https://www.cs.ucsb.edu/~rich/workload/}}. It contains 6 workload traces of a production cloud environment. We select a trace that contains 34 days of workloads of a large company with 50,000 to 100,000 employees. We partition the data into 360 parts and consider each partition as an average workload of day to create a 1-year workload data. The long-term performance of 5 IaaS providers is generated from the benchmark results published SPEC Cloud IaaS 2016 \cite{fattah2019long}. We augment the workload traces with the performance data to generate a long-term workload-performance dataset of five IaaS providers. We create the signature of each provider using the approach in \ref{sec:sig}. The experiment variables are shown in Table \ref{tab:data}. We conduct the experiments by changing the signatures randomly to create new signatures.


\begin{table}
\centering
\caption{Experiment Variables}\label{tab:data}
\begin{tabular}{|l|l|}
\hline
{\bfseries Variable Name} & {\bfseries Values}\\
\hline
Total number of simulation &  {100} \\
Total provisioning period & {360}  days \\
Trial length of each consumer & {30} days \\
Total number of IaaS performance signatures & {5} \\
Total number of Consumers &  {18} \\
Similarity thresholds & { 0.1 to 0.9 }\\
Anomaly Thresholds & {10\% to 100\%} \\
\hline
\end{tabular}
\vspace{-5mm}
\end{table}

\subsection{Evaluation and Discussion}

\begin{figure}[h]
    \centerline{
        \subfloat[]{\includegraphics[width=0.24\textwidth]{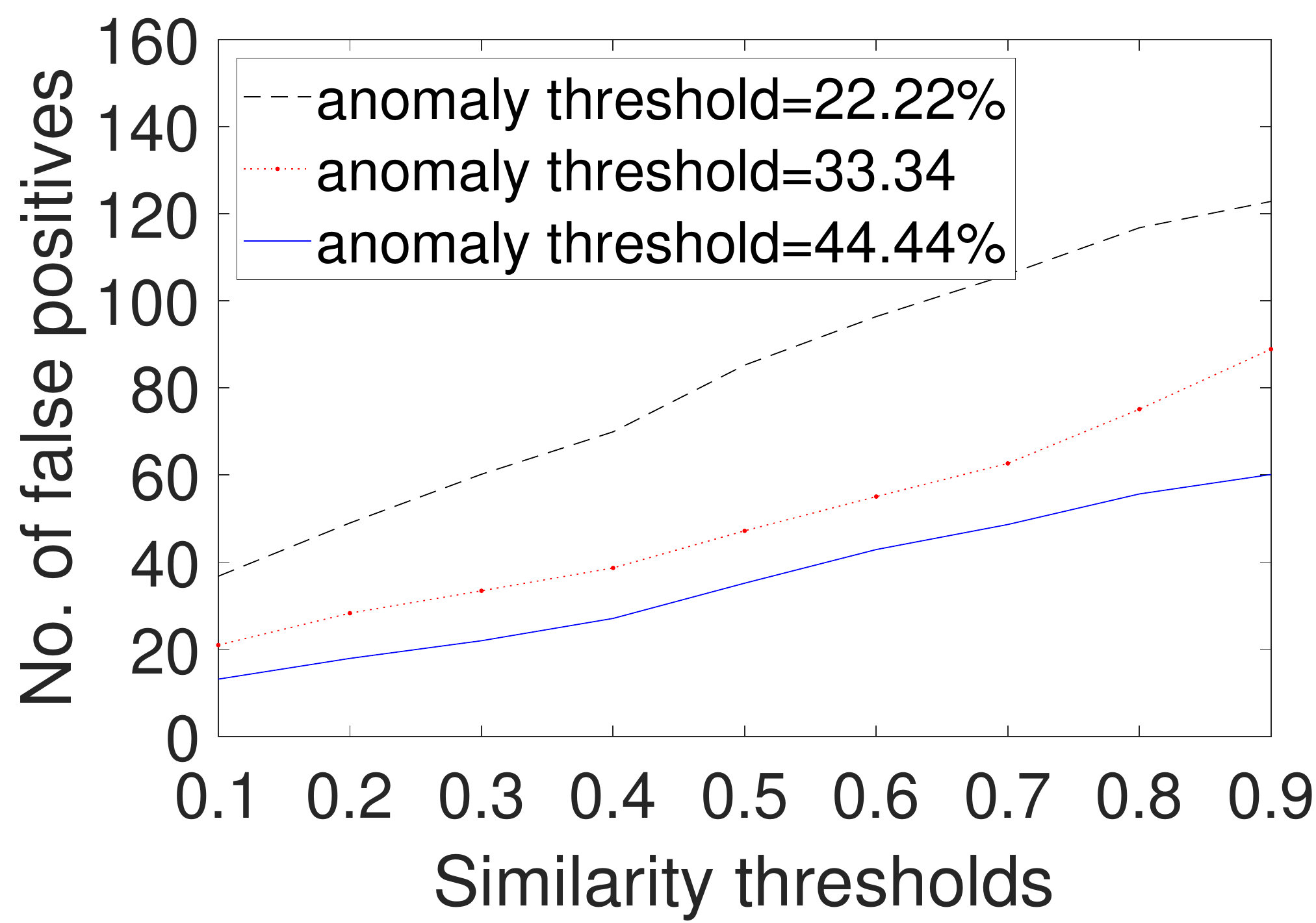} \label{fp_similarity}}
        \hfil
        \subfloat[]{\includegraphics[width=0.24\textwidth]{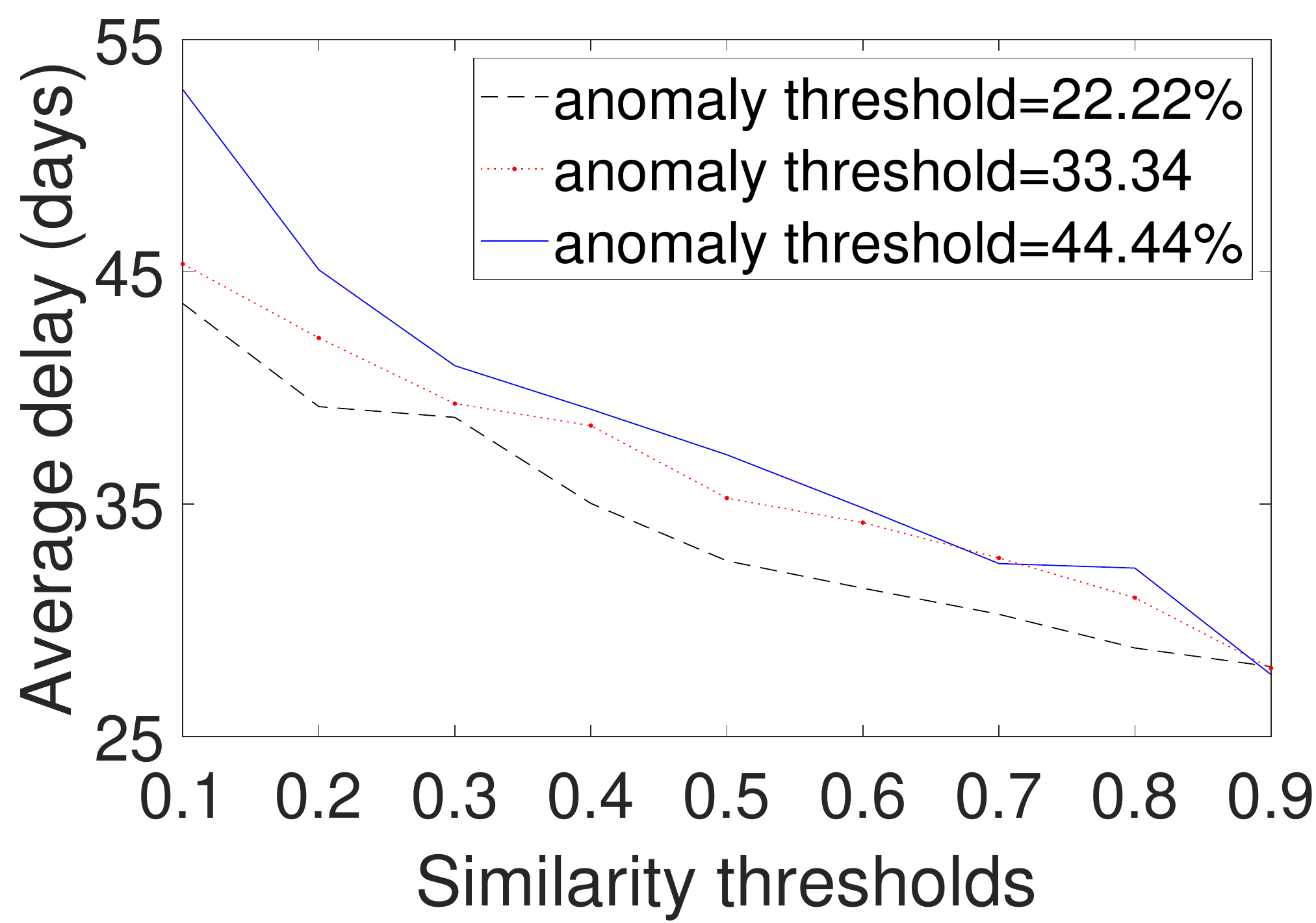} \label{delay_similarity}}
    }
 
    \caption{ Effects of different similarity thresholds in change detection in (a) number of false positives (b) average delay }
  \label{fig:exp1}
  \vspace{-4mm}
\end{figure}

The proposed approach aims at reducing the number of false positives and the change detection delay varying the similarity threshold and the anomaly threshold for the anomaly and event detection respectively. We discuss only the results of similarity measure using PCC due to the page limitation. Fig. \ref{fig:exp1}(a) shows the number of false positives that are generated before the actual change detection for the different values of the similarity thresholds. The anomaly thresholds are set from 22.22\% to 44.44\% of the total number of consumers within a given trial month. The number of false positives increases with the increase of similarity thresholds according to Fig. \ref{fig:exp1}(a). The reason is that when the similarity threshold is increased, the number of detected anomaly increases. As a result, the number of detected events also increases resulting in a high number of false positives. The number of false positives directly affects the delay in signature change detection. The average delay in change detection is illustrated in Fig. \ref{fig:exp1}(b). The average delay is reduced with the increase of similarity threshold for anomaly detection. This implies that when the number of false positives increases, the average detection delay is reduced due to the increasing number of testing.

\begin{figure}[t]
    \centerline{
        \subfloat[]{\includegraphics[width=0.24\textwidth]{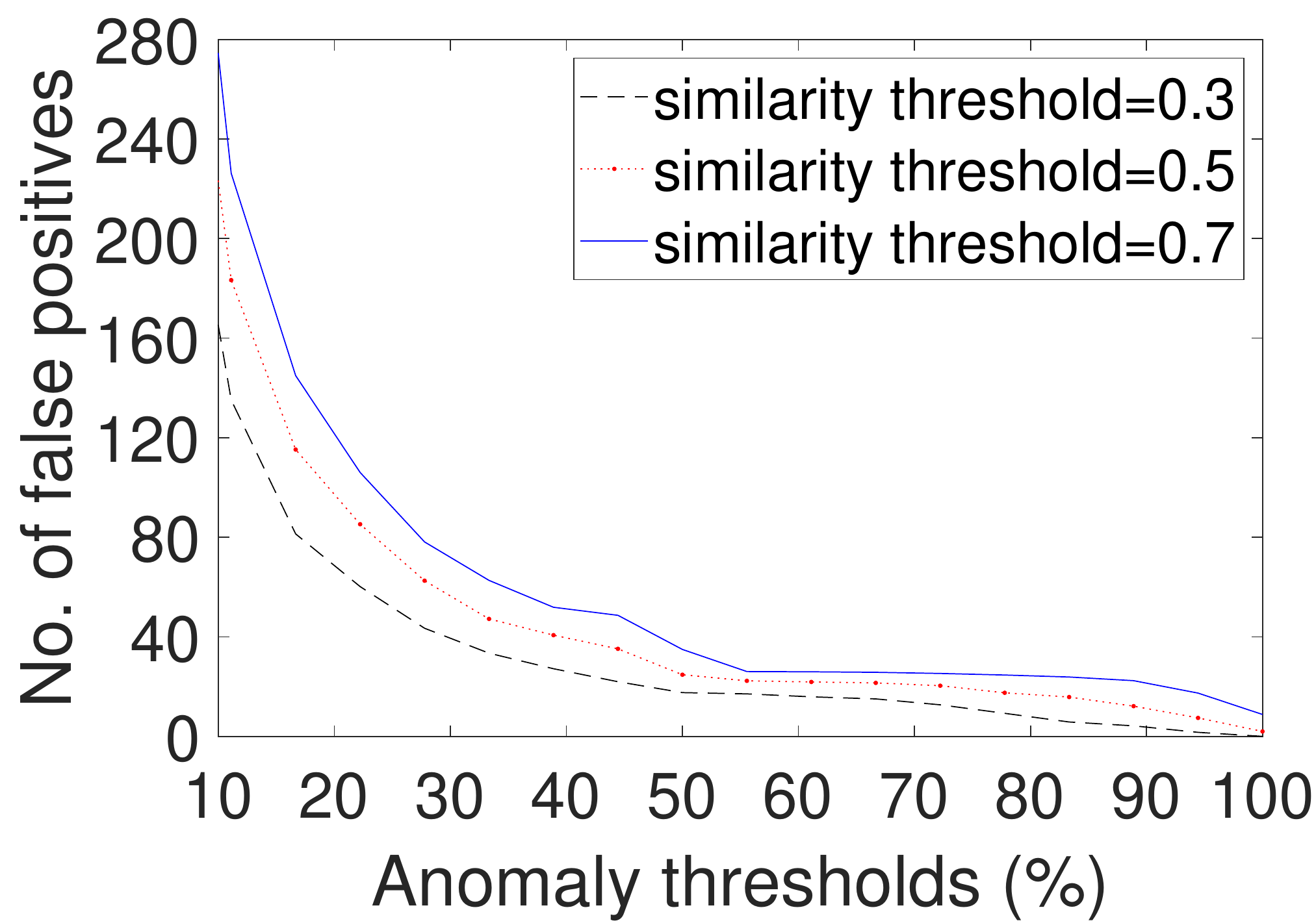} \label{fp_anomaly}}
        \hfil
        \subfloat[]{\includegraphics[width=0.24\textwidth]{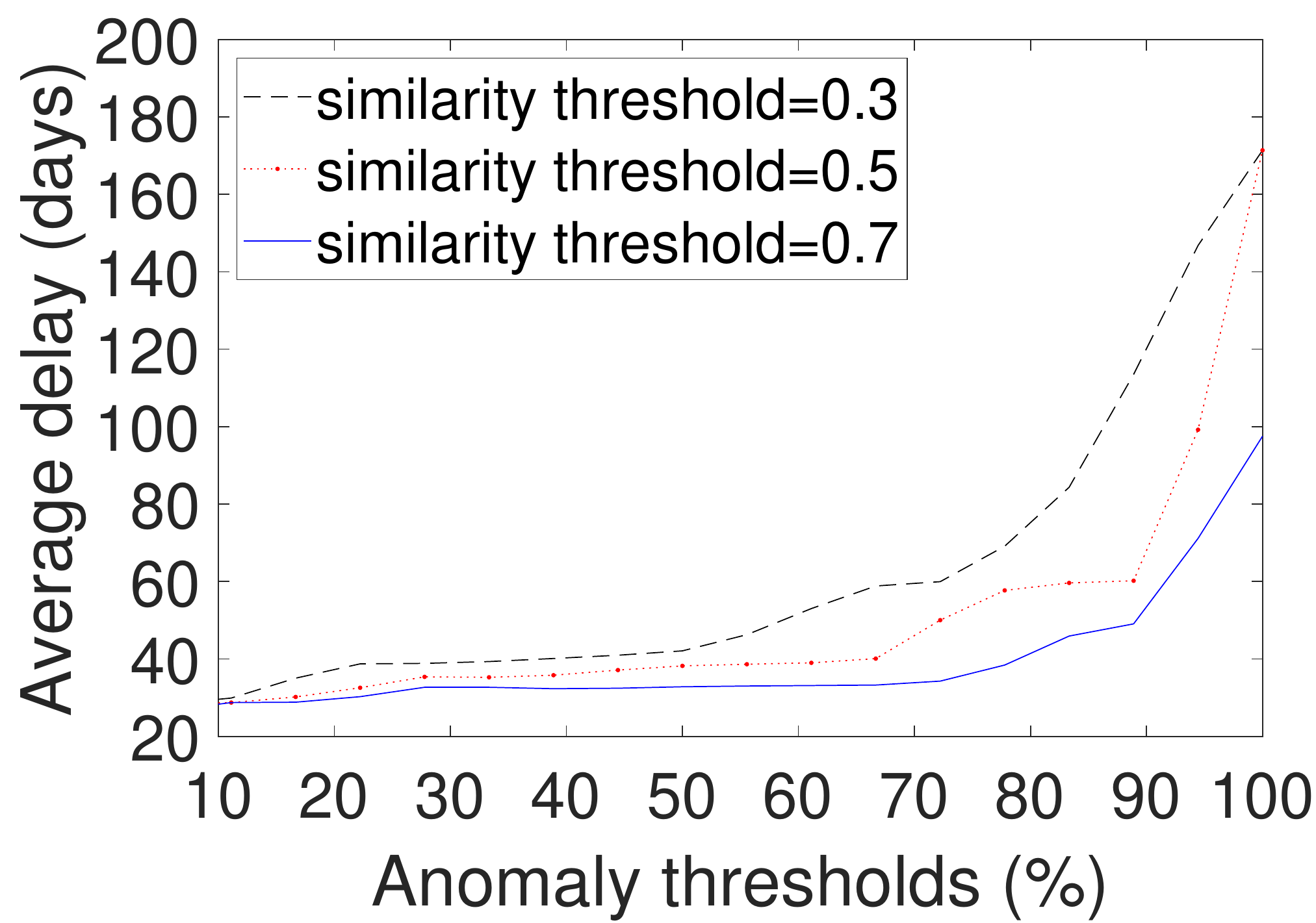} \label{delay_anomaly}}
    }
 
    \caption{ Effects of different anomaly thresholds in change detection in (a) number of false positives (b) average delay }
  \label{fig:exp2}
 \vspace{-4mm}
\end{figure}

The anomaly threshold for the event detection impacts the result in the opposite way of changing the similarity thresholds. Fig. \ref{fig:exp2}(a) and (b) illustrate the effect of changing the anomaly detection threshold on the number false positives and the detection delay respectively. The number of false positives decreases \textit{exponentially} with the increase of the anomaly threshold. For instance, when the anomaly threshold is at 100\% of the total consumer, the number of false positives becomes almost zero. The reason for such a result is that when the anomaly threshold is increased, the proposed framework accepts a higher number of performance anomalies as the normal behavior of the service. As a result, when the anomaly threshold is 100\% of the total trial users at a point of time, an event is detected only if every trial user observes anomalous performance behavior. Similarly, Fig. \ref{fig:exp2}(b) depicts that the increase in the anomaly threshold increases the average detection delay. The reason is that when the anomaly threshold is increased, the number of detected events becomes lower. When the number of events is decreased, the number of testing of signature is also decreased. This result could be inferred from the impact of anomaly thresholds on the number of false positives. Intuitively, if the number of false positives decreases, the average detection delay should increase because of the lower number of performed tests.  


\begin{figure}[b]
\vspace{-5mm}
    \centerline{
        \subfloat[]{\includegraphics[width=0.24\textwidth]{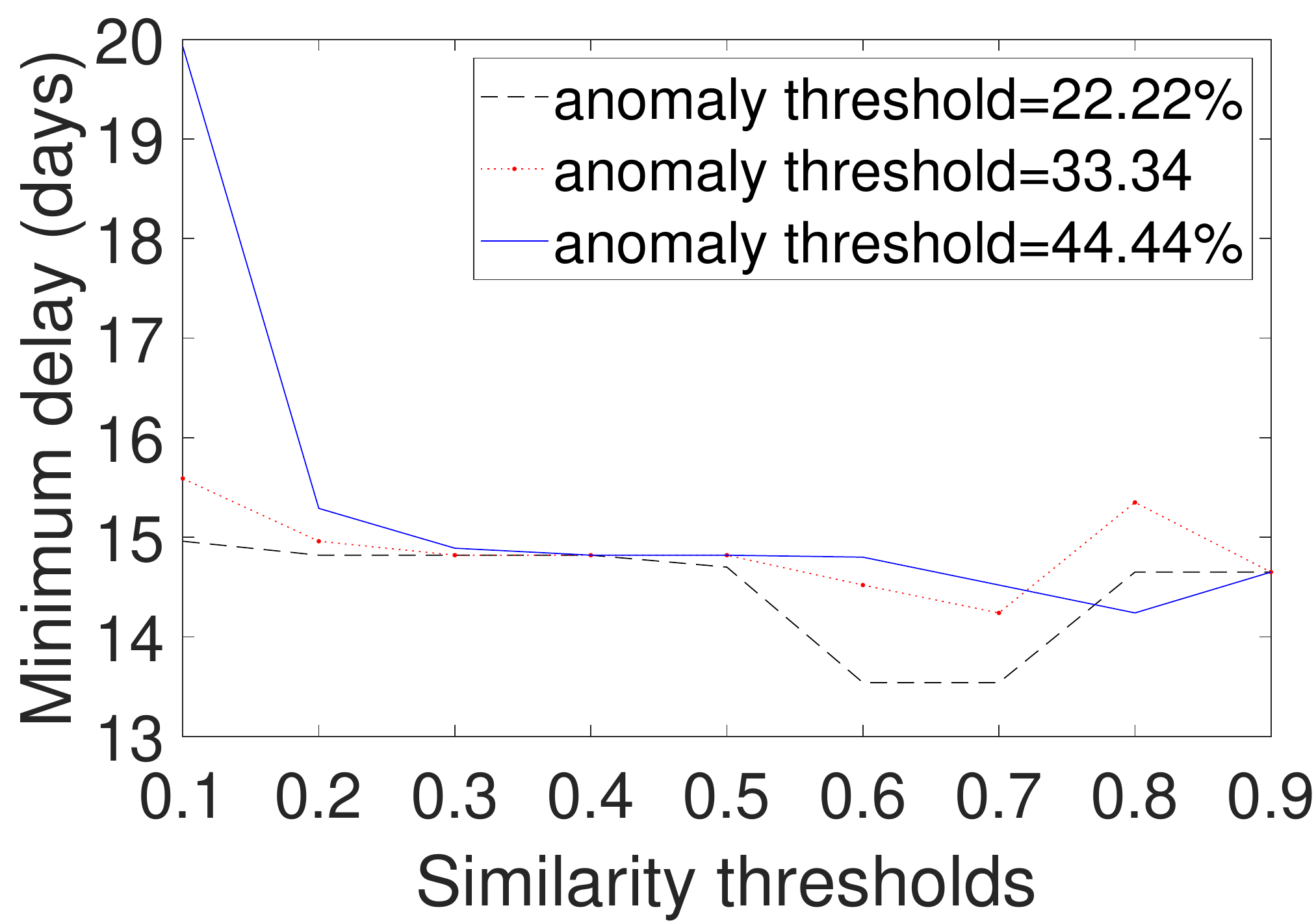} \label{min_delay_similarity}}
        \hfil
        \subfloat[]{\includegraphics[width=0.24\textwidth]{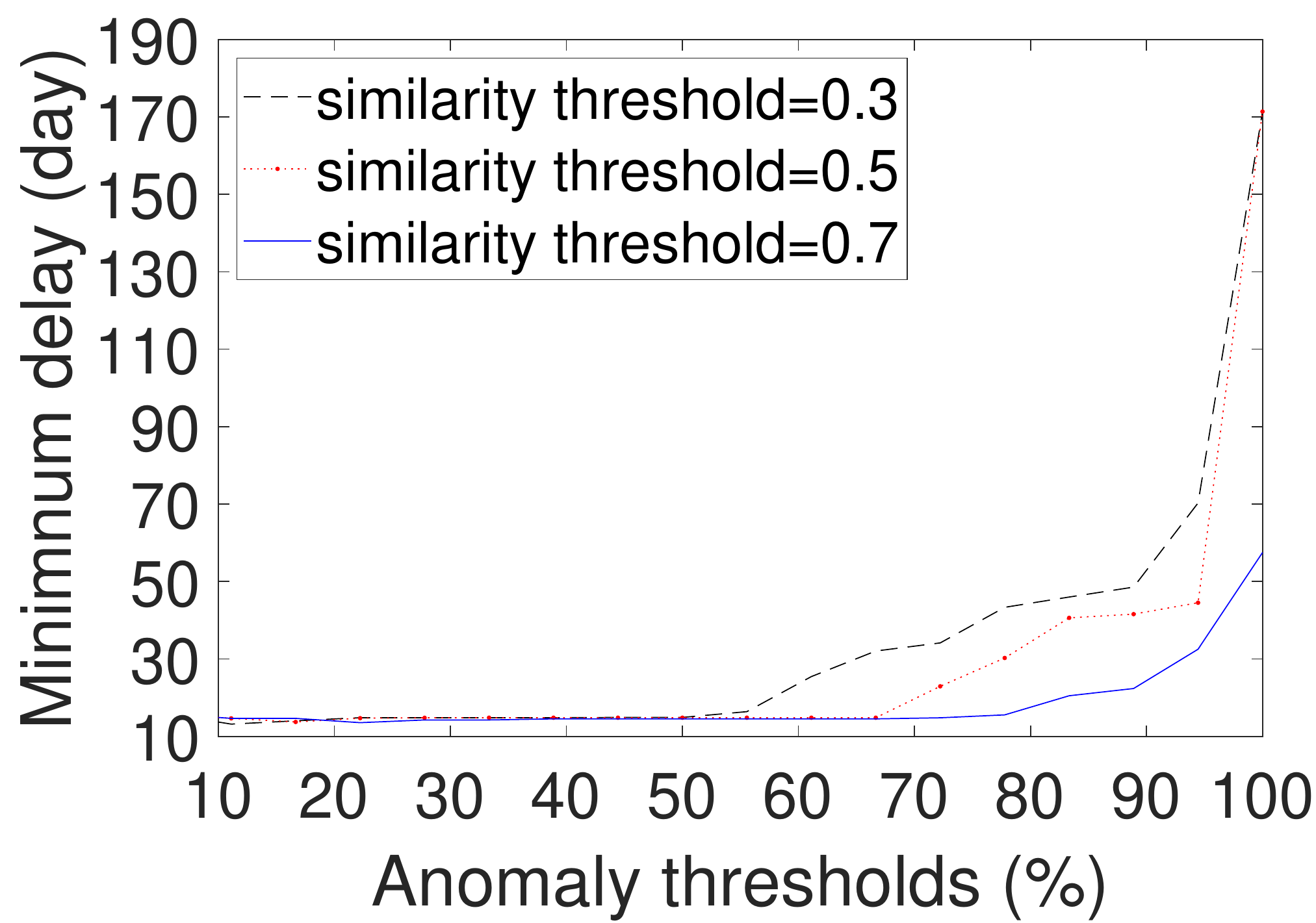} \label{min_delay_anomaly}}
    }

    \caption{Minimum delay in change detection for (a) different similarity thresholds (b) different anomaly thresholds }
  \label{fig:exp3}
\end{figure}

Fig. \ref{fig:exp1}(b) shows that the average change detection delay varies between 30 to 55 days which is reasonable given the one month trial period window. Fig. \ref{fig:exp2}(b) shows that the average detection delay varies between 15 to 180 days. The high value in the result could indicate that the proposed approach is unable to detect changes in some signatures. Fig. \ref{fig:exp3} shows that the actual delay for most of the cases is much lower. Fig. \ref{fig:exp3}(a) and (b) shows the minimum change detection delay for different similarity thresholds and anomaly thresholds respectively. The value of minimum change detection varies from 15 days to 60 days in most cases. The high value of the average change detection delay indicates that the proposed approach is unable to detect some of the changes at all. As a result, the average delay in change detection increases.

\begin{figure}
    \centerline{
        \subfloat[]{\includegraphics[width=0.24\textwidth]{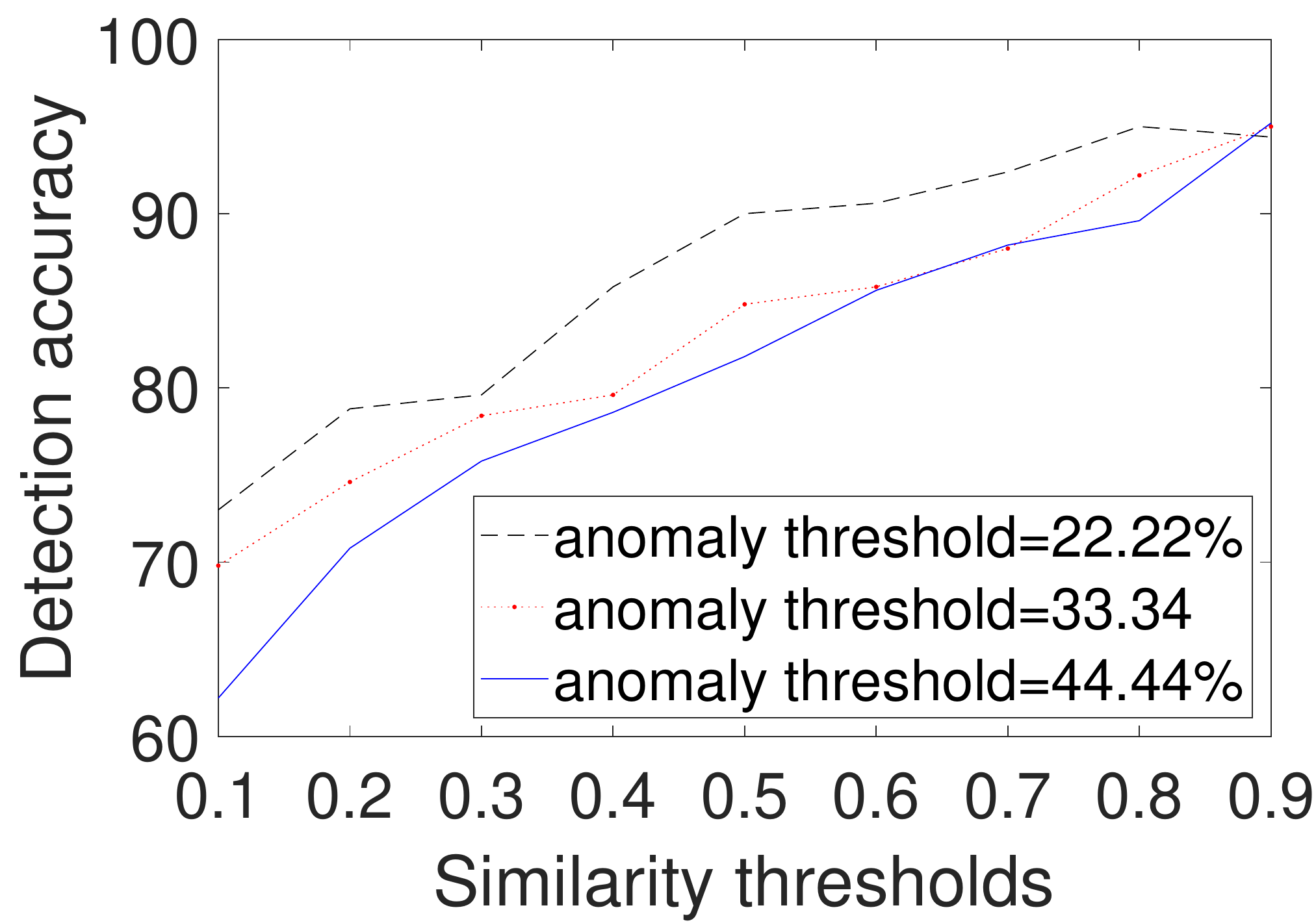} \label{accuracy_similarity}}
        \hfil
        \subfloat[]{\includegraphics[width=0.24\textwidth]{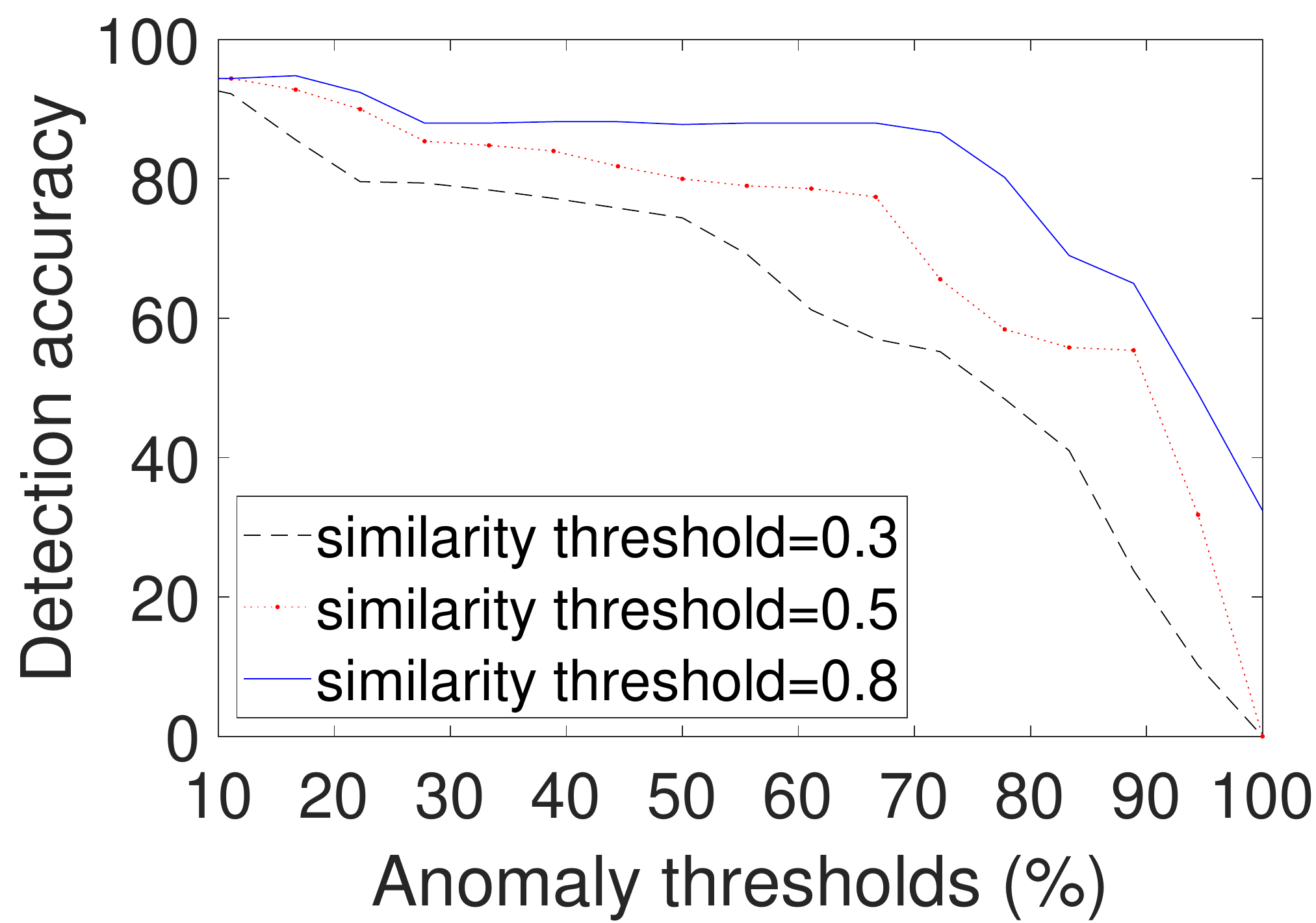} \label{accuracy_anomaly}}
    }
 
    \caption{ Accuracy of the change detection for (a) different similarity thresholds (b) different anomaly thresholds}
  \label{fig:exp4}
  \vspace{-2mm}
\end{figure}

If we wait for a long time, the proposed approach may be able to detect changes in all signature. However, waiting for an uncertain period to detect the change is unrealistic. We therefore set a \textit{time window} $T_w$. We evaluate the proposed approach in terms of its ability to detect changes in each signature within $T_w$. We set the value of $T_w$ to 60 days based on the average change detection delay as shown in Fig. \ref{fig:exp1}(b). If a change in a signature is not detected within the first 60 days of the actual change, we consider that the proposed approach is unable to detect the change for that particular signature. Fig. \ref{fig:exp4}(a) shows that the detection accuracy increases from 40\% to 95\% with the increase of the similarity threshold. This indicates that the proposed approach is able to detect the change up to 90\% of the signatures within the first 60 days when the similarity threshold is very high. It is important to note that high accuracy leads to a higher cost in terms of the number of performed tests. Fig. \ref{fig:exp4}(b) shows that the change detection accuracy decreases from 95\% to below 10\% with the increase of anomaly threshold for the event detection. This result is also consistent with the previous results. The high anomaly threshold leads to a lower number of testing which results in lower accuracy in the change detection. The accuracy results indicate that the similarity threshold and anomaly threshold are needed to be adjusted separately for each signature to improve the performance which can be performed using the proposed self-adjustment method.

\section{Related Work}

The performance variability of IaaS services is addressed in several studies \cite{iosup2014iaas,fattah2019long,leitner2016patterns}. The performance of IaaS cloud services is typically estimated for different applications based on short-trials \cite{wang2018testing,fattah2019long}. However, most of these approaches do not consider long-term IaaS performance changes. An extensive study on the variability of IaaS performance is carried out in \cite{leitner2016patterns}. The study suggests that cloud performance is a ``moving target" and requires re-evaluation periodically. A signature-based selection of IaaS cloud services is proposed in \cite{fattah2020icws}. The proposed work models the long-term performance variability of IaaS cloud services using the concept of signature. The signature of IaaS services is generated from the experience of the past trial users who share their data with a trusted third party. The trusted third party analyzes the periodic performance behavior of an IaaS service to generate its signature. However, the proposed work does not consider the changes in the signature over a long period of time \cite{fattah2020icws}. To the best of our knowledge, there is no prior work that addresses the detection of performance changes of IaaS services over a long period of time.

Performance anomaly detection is a well-studied topic in many domains including cloud computing, distributed systems, security, and software engineering. Anomaly detection strategies are classified into four major categories in \cite{ibidunmoye2015performance}, which are a) signature-based detection, b) observational detection, c) knowledge-driven detection, and d) flow and dependency analysis. We decide to choose the signature-based anomaly detection as it is a natural fit for our work. Signature-based detection doesn't require to keep historical information. As a result, we do not need to keep the record of the past trial users to detect changes in signatures.

Change detection is an important research topic that identifies abrupt changes in a process \cite{veeravalli2014quickest}. It has been applied to many domains including climate change detection, speech recognition, activity recognition, and edge detection in image processing. Existing approaches for the change detection problem are categorized as either ``offline" or ``online" methods \cite{aminikhanghahi2017survey}. Offline methods analyze the entire data set at once and find where the change had occurred. Online methods for the change detection monitor and analyze each data point as they become available from a stream or source. Online methods typically rely on the statistical properties of the process to determine the change. We identify three criteria to evaluate change point techniques: a) ability to detect changes, (b) accurately identifying the change points, and (c) the number of tests to detect changes. We apply these three criteria to evaluate the proposed ECA approach.

\section{Conclusion}

We propose a novel ECA approach to detect changes in IaaS performance which would warrant changes in the corresponding  IaaS signature. The proposed approach relies on the detection of anomalous performance behavior from the experience of free trial users to detect changes in IaaS performance. A novel anomaly-based event detection technique is proposed to determine when to trigger the re-evaluation of IaaS signatures. The experiment results show that the proposed approach is able to accurately detect changes in IaaS performance that warrant re-evaluation of the corresponding IaaS signature. Detecting changes in long-term IaaS performance is important as it will help new consumers to select the best services according to their long-term QoS requirements. In future work, we aim to conduct the experiments on a larger scale to evaluate the impact of the proposed approach in the long-term selection. 

\section{Acknowledgement}

This research was partly made possible by DP160103595 and LE180100158 grants from the Australian Research Council. The statements made herein are solely the responsibility of the authors.

\bibliographystyle{IEEEtran}
\bibliography{IEEEabrv,Main}
\end{document}